\documentclass{article}

\usepackage{arxiv}

\usepackage[utf8]{inputenc} % allow utf-8 input
\usepackage[T1]{fontenc}    % use 8-bit T1 fonts
\usepackage{hyperref}       % hyperlinks
\usepackage{url}            % simple URL typesetting
\usepackage{booktabs}       % professional-quality tables
\usepackage{amsfonts}       % blackboard math symbols
\usepackage{microtype}      % microtypography
 \usepackage{relsize}

\usepackage{amssymb}
\usepackage{amsmath}
\usepackage{mathrsfs}
\usepackage{multicol}

\usepackage{graphicx}
\usepackage{cite}

\usepackage{xcolor}
\usepackage{tabularx}

\title{A computational framework for rheologically complex thermo-visco-elastic materials}

\author{
Pietro Lenarda\\
Research Unit Multi-scale Analysis of Materials (MUSAM), IMT School for Advanced Studies Lucca, Italy. \\
              \texttt{pietro.lenarda@imtlucca.it}\\
              \And
Marco Paggi\\
Research Unit Multi-scale Analysis of Materials (MUSAM), IMT School for Advanced Studies Lucca, Italy. \\
              \texttt{marco.paggi@imtlucca.it}
}

\begin{document}
\maketitle

\begin{abstract}
Fractional calculus has been proved to be very effective in representing the visco-elastic relaxation response of materials with memory such as polymers. Moreover, in modelling the temperature dependency of the material functions in thermo-visco-elasticity, the standard time-temperature superposition principle is known to be ineffective in most of the cases (thermo-rehological complexity). In this work,
a novel finite element formulation and numerical implementation is proposed for the simulation of transient thermal analysis in thermo-rehologically complex materials. The parameters of the visco-elastic fractional constitutive law are assumed to be temperature dependent functions and an internal history variable is introduced to track the changes in temperature which are responsible for the phase transition of the material. The numerical approximation of the fractional derivative is employed via the so called Grünwald-Letnikov approximation. The proposed model is used to numerically solve some test cases related to relaxation and creep tests conducted on a real polymer (Etylene Vynil Acetate), which is used as the major encapsulant of solar cells in photovoltaics. 
\end{abstract}

\section{Introduction}
The theory of visco-elasticity deals with the description of materials that
exhibit a combination of elastic (able to recover the original undeformed state after
stress removal) and viscous (deformation-preserving after stress removal) behaviours. Quantitative description of such materials involves a
strain-stress constitutive relation that depends upon time. The classical linearized model of visco-elasticity leads to an integro-differential equation in which the elastic stress
tensor $\boldsymbol \sigma$ is a convolution product between the strain $\boldsymbol \varepsilon$, which encodes the deformation history of the material up to the current time, with an appropriate memory kernel $E(t)$ (in one dimension), representing the relaxation
mechanism:
\begin{equation}
\sigma(t)=\int^t_0 E(t-s) \dot{\varepsilon}(s) \ \mathrm{d}s.
\end{equation}
The visco-elastic constitutive response is usually assessed through experimental creep or relaxation tests. In a relaxation test, a constant strain $\varepsilon_0$ is applied quasi-statically to a uniaxial tensile bar at $t=0$. Due to visco-elasticity, the stress $\sigma$ needed to maintain the imposed strain decreases with time. The relaxation modulus of the material is defined as 
$E(t) = \sigma(t)/\varepsilon_0$, and it usually shows a power-law dependency with time for the majority of polymers. In the creep test, on the other hand, the uniaxial tensile bar is loaded by a constant stress $\sigma_0$ imposed at $t=0$. Again, the load is applied quasi-statically or in such a manner as to avoid inertia effects, and the material is assumed to have no
prior history. In this case, the strain $\varepsilon$ under the constant load increases with time and the test defines a new quantity called \emph{creep compliance} $J(t) = \varepsilon(t)/\sigma_0$.
Various mathematical models have been proposed and used to represent
the visco-elastic material functions analytically. The simplest mechanical model consists of two elements: a spring for the
elastic behavior and a dashpot for the viscous one. Spring and dashpot
elements can be combined in a variety of arrangements to produce a simulated visco-elastic response. Early models due to Maxwell and Kelvin
combine a linear spring in series or in parallel with a Newtonian damper
\cite{Bow09}:
\begin{subequations}
\begin{align}
 k \sigma+ \eta \dfrac{D \sigma}{D t}= k \eta \dfrac{D \varepsilon}{D t}  & \quad (\text{Maxwell}), \\
 \sigma =k \varepsilon+ \eta \dfrac{D \varepsilon}{D t} & \quad (\text{Kelvin-Voigt}).
\end{align}
\end{subequations}
Other basic versions include the three-parameter solid and the four-parameter fluid models. A more versatile model is obtained by connecting a
number $N$ of Maxwell elements (arms) in series and adding a spring in
parallel, leading to a Prony series relaxation function:
\begin{equation}
E(t)= E_0+ \sum^N_{i=1} E_i \textrm{exp}(-t/\tau_i),
\end{equation}
where $E_i, \tau_i$ are material properties to be determined from data. It has to be
pointed out that $E_i, \tau_i$ in the Prony series have to be fitted from real data obtained from realxation tests. This is not a straight forward
task because it involves the solution of a constrained optimization problem, as pointed out in \cite{PDS+11, EU10}. \\
Fractional calculus has been proved to be very effective in modelling
the power-law time-dependency of the relaxation behavior of polymers,
offering also an easier way to estimate the model parameters as compared to the above
rheological models \cite{MP15, MDP11, EL99, CAR11, ZIN09, DIPAOLA11, ALBA16, DIPAOLA201350, Baglieri, Sapora}. As highlighted in \cite{Bag91}, Nutting, established in 1921 that for many
polymers the relationship between stress and strain is described by an equation of the form:
\begin{equation}\label{Nut}
\sigma(t)=At^{-\alpha} \varepsilon(t), \quad 0 \leq \alpha \leq 1,
\end{equation}
where the relaxation modulus $E(t)$ in Eq. \eqref{Nut} is assumed to be a fractional
kernel function of the parametres $A > 0, 0 \leq \alpha \leq 1$ of the type:
\begin{equation}\label{Rel}
E(t) = A t^{-\alpha } /\Gamma(1 - \alpha),
\end{equation}
being $\Gamma(x)$ the Euler gamma function. In this case, the constitutive
relation \eqref{Rel} reduces to an elastic spring for $\alpha = 0$ or to a dashpot
for $\alpha = 1$, suggesting that visco-elasticity is something in between those
two limit constituive models. This is the reason why the stress-strain relation
\eqref{Rel} is usually called a spring-pot element \cite{Koe84, AS02}.
Noting that the fractional derivative of a function $f(t)$ of order $0 \leq \alpha \leq 1$ is defined as \cite{Mai10}:
\begin{equation}
D^{\alpha} f (t) =\dfrac{1}{\Gamma(1-\alpha)} \int^t_0 (t-s)^{-\alpha} \dot{f}(s) \ \mathrm{d}s,
\end{equation}
it is straightforward to recast the constitutive relation \eqref{Rel} in terms of a fractional derivative \cite{AE03, Koe84, Fab14, DLWZ15}:
\begin{equation}
\sigma(t) = AD^{\alpha} \varepsilon(t).
\end{equation}
\\
Polymers display a strong thermo-visco-elastic constitutive response, with a variation in the material properties up to
three orders of magnitude, depending on temperature \cite{MP15, EU10, PDS+11, ALBA16, DIPAOLA201350, Baglieri, Sapora}. Hence, it is a common assumption the use of the so-called time-temperature superposition principle \cite{Bow09, GM10, FL02, GN9},  which states that all the material visco-elastic functions at any temperature $T$ can directly be obtained from the
same curve, the so-called \emph{master-curve}, obtained at a base temperature $T_{\rm{ref}}$, shifted
in the time axis by a quantity $a_T$. This quantity is a material parameter and must be determined from experiments. Unfortunately, for a large class of materials, this fitting leads to poor results
showing that the time-temperature superposition principle does not always apply \cite{PDS+11, EU10}. Those materials are called thermo-rheologically
complex \cite{EL99, GM10}. \\
In the present study, a new formulation for the analysis of coupled thermo-visco-elastic material problems is proposed within the finite element method that is able to account for rheologically complex materials.
For such materials, the classical time-temperature superposition principle does not apply and a fractional calculus formulation, with parameters function of temperature, is developed. A new material function $\tau(t, T)$ is introduced,
function of time and temperature history and taking into account the
phase transition in the microstructure of the polymer due to temperature variations. Ethylene-Vynil-Acetate (EVA) is considered as a representative material showing this effect in technological applications. 
\\
In Section 2, visco-elastic constitutive equations in $3D$ are formulated, with special regard to rheologically complex materials. In Section 3, the strong and weak problems of the thermo-visco-elastic dynamics are formulated. In Section 4, the finite element formulation is derived.
Section 5 addresses a series of benchmarks numerical examples to show the capabilities of the presented model. Section 6 concludes this article, highlighting the major results and the future perspectives of the present research.  

\section{Visco-elastic constitutive equations in 3D and rheologically complex materials}
In three dimensions, the material functions characterizing completely
the response of a visco-elastic solid are the Young modulus $E(t)$, the
bulk modulus $K(t)$ and the shear modulus $G(t)$. The Young modulus
is considered to be of fractional type \eqref{Rel}, as in \cite{MP15}. Following
\cite{HS96}, the assumption of constant bulk modulus $K(t) = K$ is made.
This is because polymer materials are known to show a predominant
visco-elastic behaviour in shear deformation rather than in volumetric
expansion. The remaining shear modulus $G(t)$ is found via elastic/visco-elastic correspondence principle and inverse Laplace transform using the
Mittag-Leffler special functions \cite{SMH11}.
Let us consider a material occupying a region $R \subset \mathbb{R}^3$ in the three-dimensional space. Let $\mathbf{u}$ be the displacement field. Let 
$\boldsymbol \varepsilon= (\nabla \mathbf{u}+ \nabla \mathbf{u}^{\rm T})/2$ 
be the infinitesimal strain tensor. Assume that the material is isotropic, then decomposing the overall stress tensor $\boldsymbol \sigma$
into its deviatoric and hydrostatic parts, a visco-elastic behavior only for
the deviatoric part is herein considered. The split of the stress tensor $\boldsymbol \sigma$
reads:
\begin{equation}\label{split}
\boldsymbol{ \sigma}_{\rm d}(t)=2 \int^t_0 G(t-s) \dfrac{\partial \boldsymbol{\varepsilon}_{\rm d}(s)}{\partial t} \ \mathrm{d}s, \quad \boldsymbol{ \sigma}_{\rm v}(t)=3K \boldsymbol {\varepsilon}_{\rm v}
\end{equation}
where $\boldsymbol{ \sigma}_{\rm d}= \boldsymbol \sigma - \boldsymbol{ \sigma}_{\rm v}/3$ and
$\boldsymbol{ \varepsilon}_{\rm d}=\boldsymbol \varepsilon- \boldsymbol{ \varepsilon}_{\rm v}/3$ are the deviatoric stress and strain tensors, $G(t)$ is the shear
modulus and $K(t) = K$ is the constant bulk modulus. The visco-elastic
constitutive model \eqref{split} is of Kelvin-Voigt type.
To solve the three-dimensional visco-elastic problem, it is
necessary to know the actual mathematical expression of the three material functions $G, K$ and $E$, and their
dependency upon time. Following \cite{MP15}, a Young modulus of fractional
type is herein considered:
\begin{equation}\label{E}
E(t)=A t^{-\alpha}/\Gamma(1-\alpha), \quad 0 \leq  \alpha \leq 1.
\end{equation}
Given the espression of the Young modulus $E(t)$, an explicit time-dependency
of the shear modulus $G(t)$ is now obtained.
Let $f$ be a function and denote with $\mathcal{L}[f]$ its Laplace transform. The
$s$-multiplied Laplace transform given by $s \cdot \mathcal{L}[f]$ is denoted as $\overline{f}^*(s)$.
The elastic/visco-elastic correspondence principle \cite{Bri08} states that, in
the Laplace domain, the shear modulus is given by:
\begin{equation}\label{Lap}
\overline{G}^*(s)=\dfrac{3\overline{E}^*(s)\overline{K}^*(s)}{9\overline{K}^*(s)-\overline{E}^*(s)},
\end{equation}
then, because $K(t) = K$ and $E(t)$ is given by Eq. \eqref{E}, in the Laplace domain $K (s) = K/s$ and $E (s) = A s^{\alpha-1}$. Substituting these espressions in Eq. \eqref{Lap} and taking the inverse Laplace transform, leads to:
$$
G(t)/3=-E_{\alpha}[-9Kt^{\alpha}/A],
$$
where
$$
E_{\alpha}[x]=\sum^{\infty}_{k=0} \dfrac{x^{k}}{\Gamma(\alpha k+1)}
$$
is the Mittag-Leffler special function of order $\alpha$ \cite{SMH11, Mai10}.
Since this function does not have a closed-form expression, the following asymptotic approximation is introduced:
\begin{equation}
 E_{\alpha}[-\lambda t^{-\alpha}] \approx \begin{cases}
                                         1-\lambda \dfrac{t^{\alpha}}{\Gamma(1+\alpha)},& \quad t \to 0^+\\
                                           \dfrac{t^{-\alpha}}{\lambda \Gamma(1-\alpha)},& \quad t \to + \infty, 
                                          \end{cases}
\end{equation}
valid for any $\lambda$. From this expression, the formula found by Pipkin
in \cite{Pip12} is recovered:
\begin{equation}\label{pip}
G(t) \approx E(t)/3, \quad
t \to +\infty,
\end{equation}
which will be used in the sequel as the expression for the time-dependent
shear modulus.

\subsection{Time-temperature superposition principle and its limits}
In treating problems involving polymers, the so-called time-temperature
superposition principle is a common assumption. This principle states
that all the relaxation functions $E(t, T ), G(t, T )$ and $K(t, T )$ at any temperature $T$ can directly be obtained from the material functions at base
temperature $T_{\rm ref}$, by replacing the current time $t$ with a shift function
$a_T$, which is a material property of the material and must, in general, be
determined experimentally \cite{GN94, Chr13, KF12, ALBA16, DIPAOLA201350, Baglieri, Sapora}:
\begin{equation}\label{ttps}
E(t, T ) = E(t/a_T , T_{\rm ref} ).
\end{equation}
This is equivalent to say that, for any fixed temperature $T$, the relaxation
curve $t \mapsto E(t, T )$ is obtained from the same master-curve at a base temperature $T_{\rm ref}$, shifted along the horizontal axis by a quantity $a_T$ in a log
time scale. The shift factor $a_T$ is usually described by the WLF (Williams-Landel-Ferry) equation \cite{FW52, Voi14}
$\rm{log}(a_T)=-C_1(T-T_g)/(C2-(T-T_g))$, where $C_1, C_2$ are constants and $T_g$ is the glass transition temperature.
The previous espression is known to be valid only for $T > T g$.\\
Materials where the shifting results in a satisfactory mastercurve are
called thermo-rheologically simple. Unfortunately, this is not the case
for several materials, like for instance EVA (Ethylene-Vynil-Acetate), which is a copolymer
containing semicrystalline parts and whose microstructure changes with
temperature, udergoing several phase transitions \cite{MP15, UE11}. Those
materials are called thermo-rheologically complex \cite{Bag91}.
The relaxation curves of the Young modulus $E(t, T )$ for
EVA at different temperatures in a log-log scale are shown in
Fig.1 \cite{PDS+11, MP15}. It can be noticed that the straight lines have
different slopes in the temperature range under consideration, so that an
horizontal shifting of the material function $E(t, T )$ for different temperatures $T$ does not result in a satisfactory overlap. This result suggests
that the shift factor $a_T$ for the EVA material, as assumed by the time-temperature principle and described by Eq. \eqref{ttps} is not accurate. This
issue is confirmed looking at Fig.2 where the fitting of the shift
factor \eqref{ttps} for EVA taken from \cite{UE11} shows a very
poor result, suggesting that a refined model should be considered for materials that do not obey the time-temperature superposition principle.

\begin{figure}[h!]
 \centering
 \includegraphics[width=9cm]{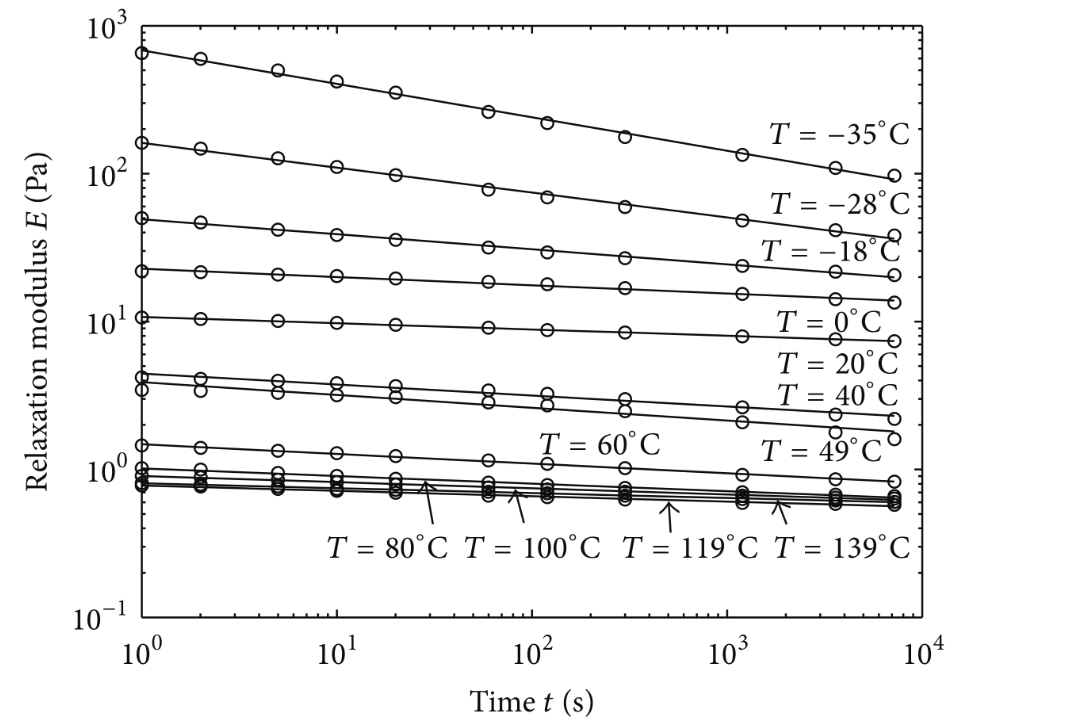}
 \caption{Experimental results of the Young modulus $E(t, T )$ over time in a log-log scale
obtained from relaxation tests at different temperatures in a range between
$-35^{\circ}C$ and $139^{\circ}C$ (dot lines) and results of the fitting according to Eq. \eqref{elastic} (solid lines).}
 \label{FIGex}
\end{figure}

\begin{figure}[h!]
 \centering
 \includegraphics[width=9cm]{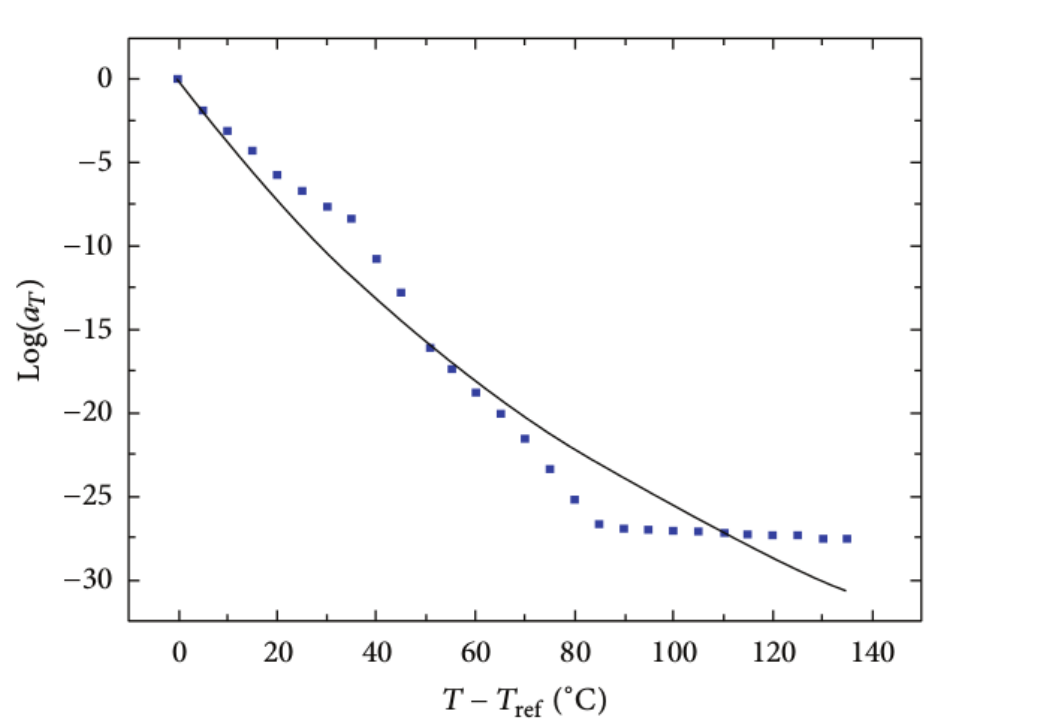}
 \caption{Fitting obtained from experimental results of the shift factor $a_T$ (continuous line) for
EVA (Etylen-Vynil-Acetate) using the WLF equation (dot line).}
 \label{FIG_1}
\end{figure}

\newpage

\subsection{A new model for thermo-rehological complexity}
Following \cite{MP15}, a temperature dependency of the Young modulus is
assumed as:
\begin{equation}\label{elastic}
E(t, T ) = A(T )t^{-\alpha(T )} / \Gamma(1 - \alpha(T )),
\end{equation}
where the material parameters $\alpha(T )$ and $A(T )$ are now temperature dependent. Functions $\alpha(T )$ and $A(T )$ can be determined from
Eq. \eqref{elastic}, fitting data from uniaxial relaxation tests conducted for different temperatures. Using the asymptotic representation of the shear modulus in Eq. \eqref{pip}, and substituing it in the first of Eqs. \eqref{split},
leads to the following integral, describing the relaxation behavior of the
material properties induced by thermal effects:
\begin{equation}\label{bad}
\dfrac{A(T)}{\Gamma(1-\alpha(T))} \int^t_0 (t-s)^{-\alpha(T)}\dfrac{\partial \boldsymbol \varepsilon_{\rm d}(s)}{\partial s} \ \rm{d}s.
\end{equation}
This espression is not convenient for applications because both $\alpha(T )$ and
$A(T )$ change continuously with temperature $T$ and time $t$ during the process, while it is better to find an espression similar, but having
constant values of $\alpha$ and $A$ as long as the internal micro-stucture of the
polymer remains the same.
To this purpose, a thermal material clock function is defined and an
espression similar to \eqref{bad}, which is suitable for applications, is introduced.
Consider an arbitrary temperature history $T(t)$, depending on the
thermo-visco-elastic process, let $\delta > 0$ be a given threshold. Let $\tau =
\tau (t, T )$ be a function which is always smaller or equal to the current time
$t$, which has the role of monitoring the temperature history inside the
material. This function is nothing but a counter taking discrete values
$0 = \tau_0, \dots, \tau_k, \dots,$ that ticks when the temperature variation exceeds the
threshold $\delta$. At the beginning of the process, $\tau = \tau_0 = 0$, and it remains
zero untill the temperature variation inside the material does not exceed
$\delta$. After that moment, the clock ticks and $\tau$ is set equal to a new value
$\tau = \tau_1$. In general, during the evolution of the process, there can be several of those temperature jumps, so that the value $\tau(t, T )$ at any time $t$ is
defined recursively as:
$$
\begin{cases}
\tau_0 & =0  \\
 \tau_k& =\rm{inf}_{\tau_{k-1} \leq t' \leq t} \{ |T(t')-T(\tau_{k-1}) | \geq \delta \}.
\end{cases}
$$
Consider the reduced time defined as: $\tau (t, T ) = \tau_k 1_{[\tau_k ,t)}(t)$, where $1_{(a,b)}$
is the indicator function of a (time) interval $(a, b)$. The function $\tau(t, T )$ becomes now a step
function, which is constant inside each interval $[\tau_{k-1} , \tau_k ]$. The modified
material functions are defined as $\tilde{A} = A(T (\tau )) = A(\tau_k ) 1_{[\tau_k ,t)}$ and $\tilde{\alpha} =\alpha(T (\tau )) =
\alpha(\tau_k ) 1_{[\tau_k ,t)}$, which are constant inside each interval $[\tau_{k-1} , \tau_k ]$.
The modified relaxation kernel $g_{A,\alpha}(t)$ is defined as:
$$
g_{A,\alpha}(t)=\tilde{A} (t-\tau)^{-\tilde{\alpha}}/\Gamma(1-\tilde{\alpha}).
$$
This physical observation suggests that the
thermo-visco-elastic relaxation process \eqref{bad} can be described by the
following fractional-thermal derivative:
\begin{equation}\label{frac-term der}
D_{A, \alpha} \boldsymbol \varepsilon_{\rm d}(t)=\int^t_0 g_{A, \alpha}(t-s) \dot{\varepsilon}_{\rm d}(s) \ \rm{d}s.
\end{equation}
Formula \eqref{frac-term der} basically says that when the variation of temperature inside the material exceeds
a given threshold $\delta$, and the thermal clock
$\tau = \tau_k$ ticks, the relaxation process is shifted backwards in time of a quantity $t - \tau_k$ and restarts with new parameters evaluated at $\alpha(T (\tau_k ))$ and
$A(T (\tau_k ))$ for all subsequent times (see Fig. 3).
This is because the material has experienced a phase transition and temperature has affected its internal microstructure.

\begin{figure}[h!]
 \centering
 \includegraphics[width=8cm]{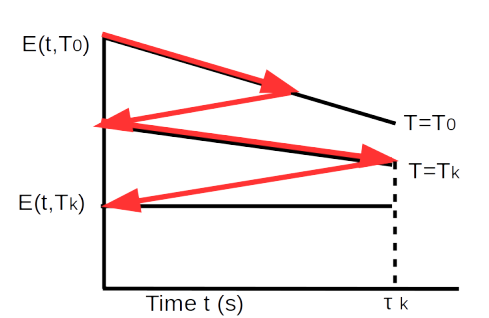} %figa11.png
 \caption{ Schematic representation of the relaxation process of the material
function $E(t, T )$ depending on the thermal history in a thermo-rheologically
complex material.}
 \label{FIG_3}
\end{figure}

Accordingly, the relaxation process for a thermo-rheologically complex material can be re-written as:
$$
\boldsymbol \sigma_{\rm d}= \dfrac{2}{3} \int^t_0 g_{A, \alpha}(t-s) \dot{\varepsilon}_{\rm d}(s) \ \mathrm{d}s= D_{A, \alpha} \boldsymbol \varepsilon_{\rm d}(t).
$$
Notice that, if the process is adiabatic, i.e. the temperature $T(t)$ remains constantly equal to the initial temperature $T_0$ during the time interval $[0, t_{\textrm{fin}} ]$, then $\tilde{\alpha} = \alpha(T_0 ) = \alpha$ and $ \tilde{A} = A(T_0 ) = A$ are constants, the
thermal clock is $\tau(t) = 0$, so that $D_{A, \alpha} \varepsilon_{\mathrm{d}} (t)$ reduces to the usual
fractional derivative:
$$
A D^{ \alpha} \varepsilon_{\mathrm{d}} (t)= \dfrac{A}{\Gamma(1-\alpha)} \int^t_0 (t-s)^{-\alpha} \dot{\varepsilon}_{\mathrm{d}} (t) \ \rm{d}s.
$$
As a concluding remark, the problem of defining an estimate for the threshold $\delta$ is adressed. This parameter defines a temperature interval in which the material properties $A$ and $\alpha$ are constants.
Looking at Fig. 2, it can be noticed that the slopes of the straight lines
are the same in different temperature intervals, i.e., in $[-35  ^{\circ}C, -28  ^{\circ} C]$,
$[-18 ^{\circ} C, 20  ^{\circ} C]$, $[40  ^{\circ} C, 139  ^{\circ}  C]$. For each of those temperature intervals, one can identify different values of $\alpha$ and $A$, and a threshold $\delta$ can be defined
accordingly, to mark the transition from a temperature interval to another.

\section{Strong and weak form of the coupled thermo-visco-elastic problem}

The deviatoric and hydrostatic parts of the stress tensor are given by:
\begin{align*}
\boldsymbol \sigma_{\textrm{d}}(t,T) &=\dfrac{2}{3} D_{A, \alpha} \boldsymbol \varepsilon_{\textrm{d}}(t), \\
\boldsymbol \sigma_{\textrm{v}}(t,T) &= 3K \boldsymbol \varepsilon_{\textrm{v}} - 3\beta (T - T_0 ),
\end{align*}
where $\beta$ is the coupling thermal stress factor and $T_0$ is the initial tempertaure inside the material. Accordingly, the overall stress tensor is given
by:
\begin{equation}\label{stress}
\boldsymbol \sigma(t,T) =\dfrac{2}{3} D_{A, \alpha} \boldsymbol \varepsilon_{\textrm{d}}(t) +
 3K \boldsymbol \varepsilon_{\textrm{v}} - 3\beta (T - T_0 ) \mathbf{I}.
\end{equation}
The balance of linear momentum takes the form:
\begin{equation}\label{linmom}
\rho \mathbf{u}_{tt} - \text{div}(\boldsymbol \sigma)=0, \quad (\mathbf{x},t) \in R \times [0, t_{\textrm{fin}}],
\end{equation}
where $\rho$ is the density of the material. Regarding the heat
conduction process, the standard Fourier law is assumed for the heat
flux $\mathbf{q}=- k \nabla T$, where $k$ is the thermal conductivity. Accordingly, the heat conduction equation reads:
\begin{equation}\label{heat}
k \nabla^2 T= \rho c T_t + \beta T_0 \dfrac{\partial \varepsilon_{\mathrm{v}}}{\partial t}, \quad (\mathbf{x},t) \in R \times [0, t_{\textrm{fin}}],
\end{equation}
where $c$ is the heat capacity of the material. Eqs. \eqref{stress}, \eqref{linmom} and
\eqref{heat} represent the system of equations for coupled linear thermo-visco-elasticity
\cite{KM12, EEB15}.

The weak form corresponding to the equation of linear momentum \eqref{linmom}
is derived by multiplying it by a virtual displacement $\mathbf{v}$ and integrating the result on the domain $R$. Applying the divergence theorem:
\begin{align*}
\int_R \rho \; \partial_{tt} \mathbf{u} \cdot \mathbf{v} \; \mathrm{d} \mathbf{x}
&+
\int_R \dfrac{2}{3} D_{A, \alpha} \boldsymbol{\varepsilon}_{\mathrm{d}}(\mathbf{u}): \boldsymbol{\varepsilon}(\mathbf{v}) \; \mathrm{d} \mathbf{x}\\
&-
\int_R \bigg( K \textrm{div}(\mathbf{u}) + \beta(T-T_0) \bigg) \textrm{div}(\mathbf{v}) \; \mathrm{d} \mathbf{x}=
\int_{\Gamma} \overline{\mathbf{t}} \cdot \mathbf{v} \; \mathrm{d} \Gamma.
\end{align*}
Analogously, the weak form
corresponding to the heat conduction partial differential equation \eqref{heat}
is obtained by multiplying it for a test function $s$ and integrating the
result over $R$
$$
\int_R k \nabla T \cdot \nabla s \; \mathrm{d} \mathbf{x}+
\int_R \rho c \partial_t T s \; \mathrm{d} \mathbf{x}+
\int_R \beta T_0 \;  \textrm{div}(\partial_t \mathbf{u}) s \; \mathrm{d} \mathbf{x}+
\int_{\Gamma} \overline{q}_n s \; \mathrm{d} \Gamma=0
$$
where the $q_n$ is the imposed normal heat flux $\nabla T \cdot \mathbf{n} = \overline{q}_n$ imposed on
the Neumann part of the domain. \\

\section{Finite element formulation}
Regarding the finite element formulation, consider a decomposition of the domain $R$ into a finite number of elements and let $\{ \boldsymbol{\Phi}^u_k \}^N_{k=1}$ and $\{ \Phi^T_k \}^N_{k=1}$ basis of shape functions. At the element level, the displacement and temperature are interpolated as:
$$
\mathbf{u}(\mathbf{x},t)=\sum^N_{k=1} \boldsymbol{\Phi}^u_k(\mathbf{x}) \mathbf{U}_k(t), \quad
T(\mathbf{x},t)=\sum^N_{k=1} \Phi^T_k(\mathbf{x}) \mathbf{T}_k(t).
$$
Discretizing the time interval $[0, t_{\rm fin} ]$ into $0 = t_0 \leq \dots  \leq t_n \leq t_N = t_{\rm fin}$
where $t^{n+1} = t^n + \Delta t$ the system of equations is:
\begin{equation}\label{disp}
\mathbf{M} \dfrac{D^2 }{D t^2} \mathbf{U}^{n+1} + \mathbf{C}^{u,T} \mathbf{T}^{n+1}+
\mathbf{G} D_{A, \alpha} \mathbf{U}^{n+1}+\mathbf{K}^u  \mathbf{U}^{n+1} = \mathbf{F},
\end{equation}
and 
\begin{equation}
 \mathbf{C} \dfrac{D }{D t} \mathbf{T}^{n+1}+ \mathbf{C}^{T,u} \dfrac{D }{D t} \mathbf{U}^{n+1}+
\mathbf{K}^T  \mathbf{T}^{n+1}+\mathbf{Q}  = \mathbf{0},
\end{equation}
where the global matrices and vectors result from the usual assembling
of matrices at the element level. In particular, $\mathbf{M}$ is the mass matrix, $\mathbf{C}^{u,T}=(\mathbf{C}^{T,u})^{ \rm T}$ is the coupling thermo-mechanical matrix, $\mathbf{G}$ and $\mathbf{K}^u$ are the shear and bulk matrices and $\mathbf{F}$ is the load vector, $\mathbf{C}$ is the termal dumping matrix, $\mathbf{K}^T$ is the thermal stiffnes matrix and $\mathbf{Q}$ is the thermal loading vector.
\\
The problem of approximating the fractional derivative $D_{A, \alpha} \mathbf{U}^{n+1}$ in Eq. \eqref{disp}
is now adressed. Let $f(t)$ be a function defined in an interval $[0, t_{\rm fin} ]$ and let
$0 = t^0 \leq \dots \leq t^n \leq t^N = t_{\rm fin}$ be a partition of $[0, t_{\rm fin} ]$, where $t^{n+1} = t^n + \Delta t$.
The Grünwald-Letnikov approximation of the fractional derivative $D^{\alpha} f (t^{n+1} )$ of order $0 \leq \alpha \leq 1$ of a $f$ function (see also
\cite{AS02}, \cite{RS11}, \cite{Mai10}, \cite{MDP11}, \cite{WS01}) reads:

\begin{equation}\label{frac}
D^{\alpha}f(t)=(\Delta t)^{-\alpha} \sum^n_{j=0} c_{j+1}(\alpha) f(t^{n+1-j})
= (\Delta t)^{-\alpha}
\left( f(t^1), \dots, f(t^{n+1}) \right) \begin{pmatrix}
                                c_{n+1}(\alpha) \\
                                \vdots \\
                                 c_1(\alpha) 
                               \end{pmatrix}  ,
\end{equation}
where the coefficients $c_j(\alpha)$ are defined by the recursive formula:
\begin{equation}\label{ci}
c_j(\alpha)=
\begin{cases}
 \dfrac{(j-1-\alpha)}{j} c_{j-1}(\alpha)  & \quad j>1, \\
 1 & \quad j=1 .
\end{cases}
\end{equation}
Coefficients in Eq. \eqref{ci} have the properties that $c_j (\alpha) < c_{j+1} (\alpha) < 0$ for $j > 1$
and $\lim_{j \to +\infty} c_j (\alpha) = 0$. Notice that the dimensions of vectors in \eqref{frac}
are increasing with $n$. Each value $f(t^n )$ up to time $t^{n+1}$ is contribuing
to the final value of $D^{\alpha}f (t^{n+1})$, but the influence of the coefficients is
weaker in the past rather than in the present and the initial value $f (t^0 )$ is
multiplied by $c_{n+1} (\alpha)$, which is tending to zero as $n$ grows. This property
of the Grünwald-Letnikov approximation is called memory effect.
\\
With this set up, the fractional-thermal derivative $D_{A ,\alpha }\mathbf{U}^{n+1}$ is
approximated as:
$$
D_{A, \alpha} \mathbf{U}^{n+1}=A(T^m) (\Delta t)^{-\alpha(T^m)} \left( \mathbf{U}^{n+1} + F^m_T \mathbf{U}^{n} \right),
$$
where $m \leq n$ is a discrete history variable depending on the current time $t^n$ and
temperature $T^n$, taking values $m_0 < m_1 < \dots < m_k$, defined recursively
as $m_0 = 0$ and:
$$
m_k=\begin{cases}
 m_{k-1}  \quad & \text{if} \ \vert T^n -T^{n-1} \vert < \delta ,\\ 
 n \quad  & \text{otherwhise}
\end{cases}
$$
and the operator $F^m_T \mathbf{U}$, collecting the displacement history up to time $n$, is given by:
\begin{equation}\label{forcing}
F^m_T \mathbf{U}=
\begin{cases}
 \mathlarger{\sum^n_{j=m}} c_{n+2-j}(\alpha(T^m)) \mathbf{U}^j \quad & \text{if} \ m<n \\ 
 \mathbf{0} \quad  & \text{if} \ m=n \ .
\end{cases}
\end{equation}.

Notice that the last term in the sum \eqref{forcing} is $c_2 (\alpha(T^m ))\mathbf{U}^n$, because the coefficient
$c_1(\alpha(T^m )) = 1$ is associated to the unknown vector $\mathbf{U}^{n+1}$. When the
process begins, the sum starts to pile up following the Günwald approximation with material functions $A(T^0 )$ and $\alpha(T^0 )$ evaluated at the starting
temperature $T^0$, untill the condition $\vert T^n - T^{n-1} \vert > \delta$ is verified. After
that moment, the history variable $m$ is set equal to $n$ and the process restarts with new
material parameters $A(T^n )$ and $\alpha(T^n )$.

When the process is adiabatic, i.e. the temperature is constant during
time, then $m = 0$ for all $t^n$ and $A = A(T_0 )$, $\alpha = \alpha(T_0 )$, so that the approximation of the thermal-fractional derivative $D_{A , \alpha }\mathbf{U}^{n+1}$ reduces to
$AD^{\alpha} \mathbf{U}^{n+1}$, which is exactly the usual Günwald-Letnikov approximation:
$$
AD^{\alpha} \mathbf{U}^{n+1}= A(\Delta t)^{-\alpha} \left[ \mathbf{U}^{1} |  \dots  |  \mathbf{U}^{n+1} \right]  \begin{pmatrix}    c_{n+1}(\alpha) \\
                                \vdots \\
                                 c_1(\alpha)
\end{pmatrix}.
$$
This implies that the displacement history must be stored in a matrix whose number of columnsis progressively increasing and represents the memory of the
material. In numerical treatment, the number of columns of the matrix is given by the number of timesteps of the problem and the matrix is initialized to be equal to the zero matrix, then each column is replaced with the displacement solution at the previous timestep.

\section{Numerical examples}
In this section, several experiments in one and two dimensions are considered to test the new thermo-visco-elastic model which has been introduced in this work.

\subsection{Free vibrations of a visco-elastic 1D rod}
\begin{figure}[h!]
 \centering
 \includegraphics[width=13.5cm]{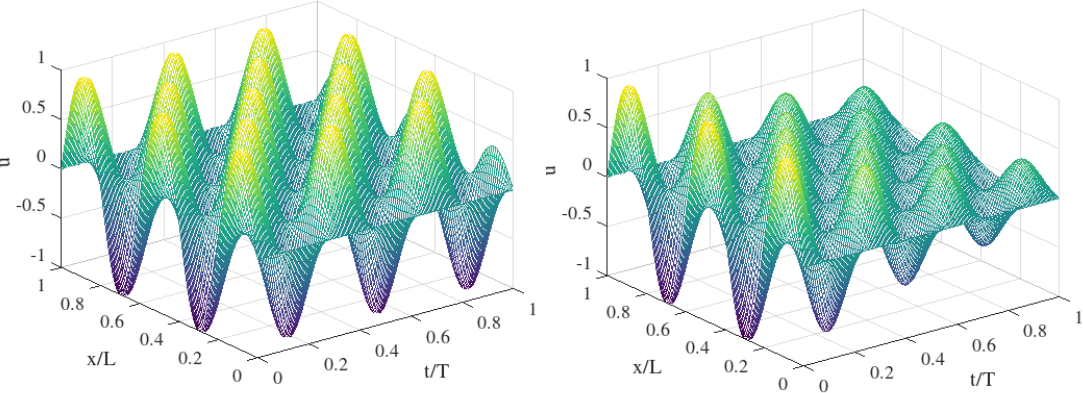} %figa11.png
 \caption{Spatio-temporal plot of the numerical solution of the adimensionalized fractional visco-elastic wave equation for different exponents of the fractional derivative $\alpha=0$ (linear elastic case) and $\alpha=0.2$. }
 \label{alpha0_01_02_03}
\end{figure}

\begin{figure}[h!]
 \centering
 \includegraphics[width=8cm]{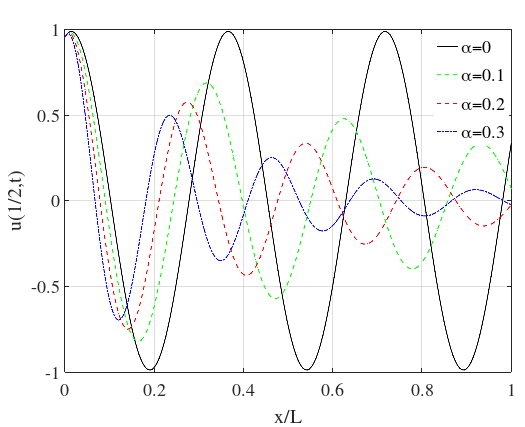} %figa11.png
 \caption{ Temporal evolution of the numerical solution of the vertical displacement $u(1/2, t)$ in the $1D$ fractional visco-elastic wave equation for different exponents of the fractional derivative $\alpha=0$ (solid line) and $\alpha = 0.1, 0.2$ and $0.3$ (dashed line). }
 \label{rod}
\end{figure}

In this example, the problem of finding the vertical displacement $u(x)$ of a one
dimensional visco-elastic vibrating rod of lenght $L$ clamped at its ends $u(0) = u(L) =
0$ is considered, being subjected to an initial sinusoidal prerturbation $u_0 (x) = \text{sin} (\pi x)$ at
time $t = 0$ and then left free to its own vibration without any external force or traction imposed. The motion of the rod is governed by the
following equation which has to be solved in space $0 \leq x \leq L$ and time $0 \leq t \leq T_f$ :
\begin{equation}\label{1d}
\rho u_{tt}- \dfrac{\partial}{\partial x} \left( \dfrac{A}{\Gamma(1-\alpha)} \int^t_0 (t-s)^{-\alpha} u_{xt}(s) \ \textrm{d}s \right)=0,
\end{equation}
where $\rho$ is the linear density of the rod and $A > 0$ and $0 \leq \alpha \leq 1$ are visco-elastic material parameters. Notice that for $\alpha = 0$ this problem reduces
to the usual wave equation in one dimension for a linear elastic rod:
$$
\rho u_{tt} - A u_{xx} = 0.
$$
In the limit case $\alpha = 0$, the mechanical energy of the system is given by:
$$
E(t)=\dfrac{1}{2} \int^L_0 \left[ \rho \left( \dfrac{\partial u}{\partial t} \right)^2 + \left( \dfrac{\partial u}{\partial x} \right)^2  \right] \ \mathrm{d} x.
$$
A global property of the solution of this problem when $\alpha = 0$ is that
$\dot{E}(t) = 0$, i.e. the mechanical energy is conserved. Every numerical
method applied to solve this problem must be able to represent
this global property of the solution in the linear elastic limit $\alpha = 0$. Let
$h$ be the spatial mesh size, $N$ the number of vertices of
the mesh and $\{\Phi_a (x) \}^N_{a=1}$ a basis of linear triangular lagrangian shape
functions, then the global $N \times N$ mass and stiffness martices $\mathbf{M}$, $\mathbf{K}$
are explicitely given by the tridiagonal matrices:
$$
\mathbf{M}= \rho 
\begin{pmatrix}
 \dfrac{2h}{3} & \dfrac{h}{6} & \dots & 0 \\
 \dfrac{h}{6}  & \dfrac{2h}{3} & \dots & 0 \\
 \vdots & \vdots & \ddots & \vdots \\
 0 & 0  & \dots & \dfrac{2h}{3}
\end{pmatrix}, \quad 
\mathbf{K}= A 
\begin{pmatrix}
 \dfrac{2}{h} & -\dfrac{1}{h} & \dots & 0 \\
 -\dfrac{1}{h}  & \dfrac{2}{h} & \dots & 0 \\
 \vdots & \vdots & \ddots & \vdots \\
 0 & 0  & \dots & \dfrac{2}{h}
\end{pmatrix}.
$$
The nodal displacement vector is given by
$\mathbf{U} =(\mathbf{U}^1, \dots, \mathbf{U}^N)^{\rm{T}}$ and the differential system arising from the FE discretization is:
$$
\mathbf{M} \dfrac{D^2 \mathbf{U}}{D t^2} + \mathbf{K} D^{\alpha} \mathbf{U}= \mathbf{0}.
$$
In Fig. \ref{alpha0_01_02_03} are shown the evolution in space and time of the numerical
solution of the adimensionalized Eq. \eqref{1d} for two
different values $\alpha = 0$ and $\alpha = 0.2$ using a central difference time integration scheme. The parameters used for the simulation are $N_s=100$ spatial nodes and $N_t=200$ temporal nodes. Fig. \ref{rod} shows the evolution
over time of the solution $u(1/2, t)$ for different values of the fractional exponent $\alpha=\{ 0, 0.1, 0.2, 0.3 \}$. The
Grünwald-Letnikov approximation of the fractional derivative $D^{\alpha} \mathbf{U}$ is
used, this leads to the formation of a (pseudo-load or forcing) vector $\mathbf{F}^n$ which represents the dislacement history up to the current time. This residual load
vector has the effect of dissipating the mechanical energy of the system,
according to the memory effect of the material.

\newpage
\subsection{Temperature behavior of the relaxation modulus }
\begin{figure}[h!]
 \centering
 \includegraphics[width=8cm]{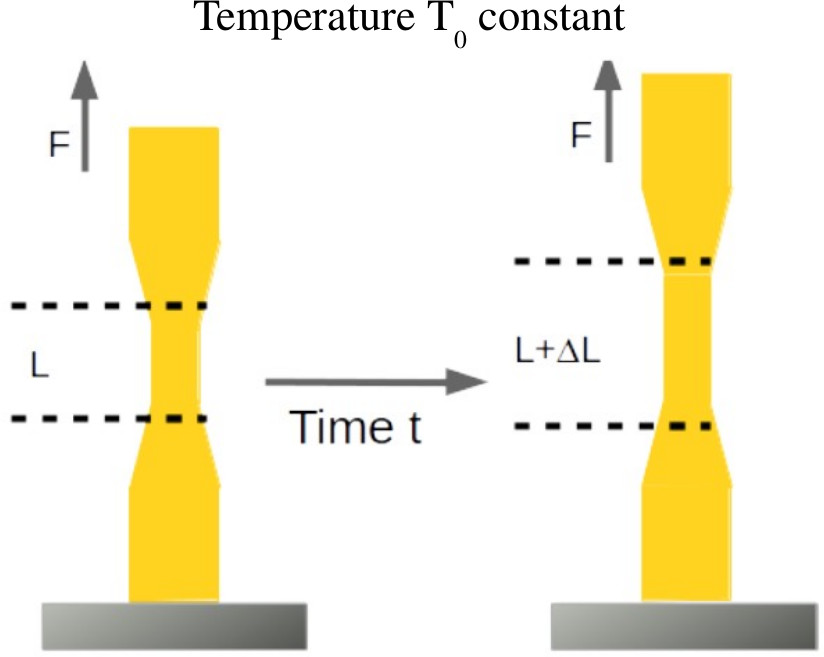} %figa11.png
 \caption{ Schematic representation of a creep test.}
 \label{es3}
\end{figure}

\begin{figure}[h!]
 \centering
\includegraphics[width=13.5cm]{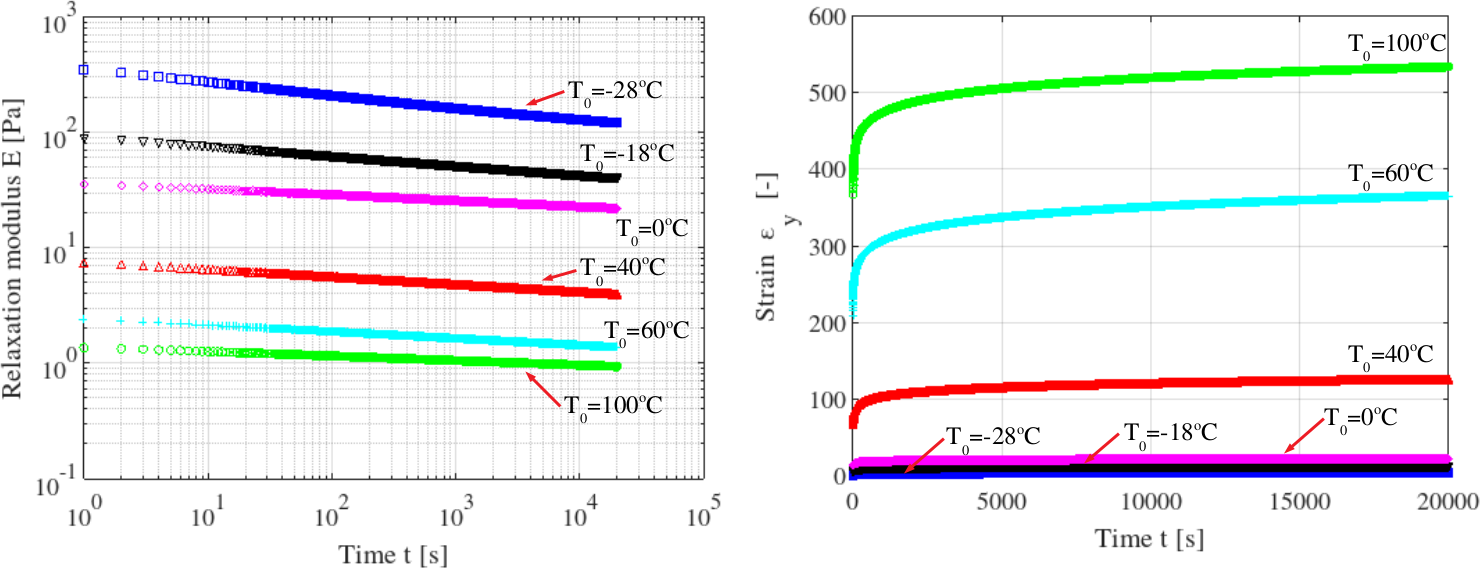} 
 \caption{Numerical results obtained from the simulations of the relaxation modulus $E$ (left) and vertical component of the strain $\varepsilon_{y}$ versus time for EVA specimen at
different temperatures obtained for fixed temperatures $-28 \; ^{\circ} \rm C, -18\; ^{\circ} \rm C, 0\; ^{\circ} \rm C, 40\; ^{\circ} \rm C, 60\; ^{\circ} \rm C, 100\; ^{\circ} \rm C$.}
 \label{rela}
\end{figure}

One of the fundamental tests used to characterize the visco-elastic
time-dependent behavior of a polymer is the creep (and creep recovery) test. In a creep test,
a constant stress $\sigma_0$ is applied quasi-statically to a uniaxial tensile bar at
zero time and held constant, as shown in the schematic Fig. \ref{es3}.
The strain, under the constant load, increases with time up to a constant value $\varepsilon_0$. A specimen of size $H \times L$ is clamped on the bottom side
and a traction $\mathbf{F}$ is applied on the top of the beam so that
$\sigma_0 = || \mathbf{F} ||$. The inertia of the beam is neglected. 
The variation of the Young's modulus with temperature can be determined from relaxation and creep tests
conducted at different constant temperatures. Material parameters $\alpha(T)$ and
$A(T)$ have been experimentally found for a polymeric material, namely Etlylene Vynil Acetate
at different temperatures and are taken from \cite{MP15} as they are reported in Table 1.
A specimen of lateral size $L = 0.02 \ \rm m$ and
vertical size $H = 0.08 \ \rm m$ is subjected to a constant traction $\sigma_0 = 500 \ \rm N$ on
the top size. During the process, the temperature is held
constant $T=T_0$. The Young's modulus is evaluated as:
$$ 
E(t)=\sigma_0/ \varepsilon_y(t).
$$
Since this process is adiabatic, the Grünwald-Letnikov approximation of the fractional derivative is employed
with constant coefficients $\alpha$ and $A$. 
Numerical results of the relaxation curves as a function of time $E(t, T)$ obtained for different constant temperatures are reported in Fig. \ref{rela} in a log log scale
and are in good agreement with the experimental ones in Fig. \ref{FIGex}. Input parameters for the simulation are the Poisson coefficient $\nu=0.29$, the Young modulus $E=3500 \ \rm MPa$, the bulk modulus $K=E/3(1-2 \nu)$ and the constant traction imposed at the upper side of the specimen is $\sigma_0=500 \ \rm N$. The input values of temperatures and corresponding fractional parameters obtained from a best fitting in \cite{MP15} are reported in Table \ref{Table_D}. It is worth noting that, as expected experimentally there is a change in the slope of the curves due to different temperatures, which is a thermo-rehologically complex behaviour of the material. In particular the elastic modulus $E(t,T)$ experiences a phase transition due to the variation of temperature which the proposed model is able to predict.

%%%%%%%%%
\begin{table*}[h!] 
\centering 
\caption{Identified parameters from \cite{MP15} $\alpha$ and $A$ used in tne numerical simulations of uniaxial
relaxation tests at different temperatures.} 
\begin{tabular}{c | c | c   }
\hline  \noalign{\smallskip} \hline\noalign{\smallskip} 
\multicolumn{1}{c}{$T_0$} &
\multicolumn{1}{c}{$\alpha$ } &
\multicolumn{1}{c}{$A \; [ \rm Pa \; s^{\alpha}]$ } 

\smallskip
\\
\hline 
 $-28 \; ^{\circ} \rm C$  & 0.16810 & 182.7   \\
 $-18 \; ^{\circ} \rm C$  & 0.10150 & 52.63   \\
 $0   \; ^{\circ} \rm C$  & 0.05566 & 23.55   \\
 $40  \; ^{\circ} \rm C$  & 0.07417 & 4.668   \\
 $60  \; ^{\circ} \rm C$  & 0.06542 & 1.544   \\
 $100 \; ^{\circ} \rm C$  & 0.04179 & 0.9276  \\

\hline  \noalign{\smallskip}\hline\noalign{\smallskip}
\end{tabular}\label{Table_D}
\end{table*}

\newpage
\subsection{The role of truncation in the Grünwald-Letnikov fractional derivative}
In this numerical example, the role of truncation in the approximation of the fractional derivative using the Grünwald-Letnikov is investigated. The relaxation test is the same as in the previous example conducted for constant temperatures $T_0=-28 ^{\circ} \rm C, -18 ^{\circ} \rm C, 0 ^{\circ} \rm C$ with a timestep $\Delta t=0.1 \ \rm s$ and $T=80000 \ \rm s$, which is enough to ensure the stationary convergence of the relaxation process. A parametric study is made by changing the memory horizon $T_{\rm h}$ to assess the effect of truncation of the fractional derivative. 

\begin{figure}[h!]
 \centering
\includegraphics[width=9cm]{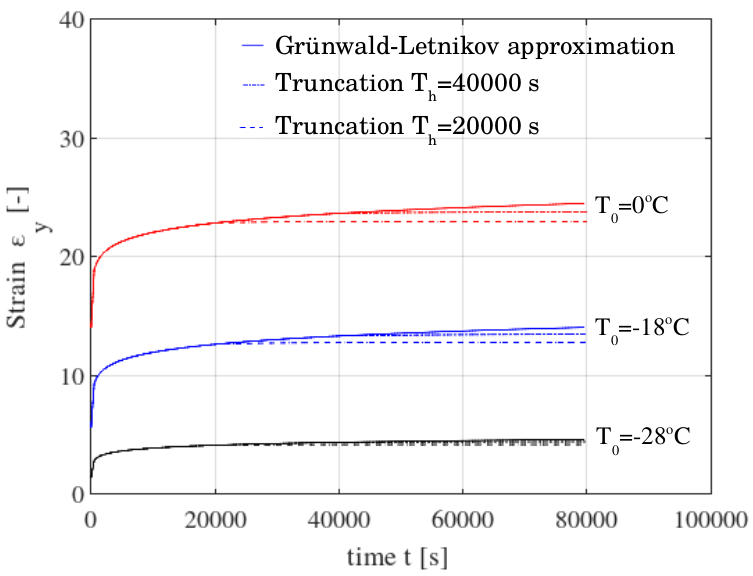} 
 \caption{Solution of the relaxation problem for temperatures $T_0=-28 ^{\circ} \rm C, -18 ^{\circ} \rm C, 0 ^{\circ} \rm C$ obtained with the full Grünwald-Letnikov approximation (solid lines) and with a truncation of the displacement history of $T_{\rm h}=40000 \ \rm s$ (dot dashed lines) and $T_{\rm h}=20000 \ \rm s$ (dashed lines) 
 .}
 \label{trunc}
\end{figure}

The memory horizon is varied as $T_{\rm h}=20000 \ \rm s, 40000 \ \rm s$. The first part of the process up to time $T_{\rm h}$ follows the Grünwald-Letnikov method. After that, the visco-elastic dynamics at current time $t$ is calculated by using the hystory from time $t-T_{\rm h}$ up to time $t$, instead of considering all the values of the displacement history from $t=0$ to $t$ as follows: 
$$
A D^{\alpha} \mathbf{U}^{n+1}= A(\Delta t)^{-\alpha} \left[ \mathbf{U}^{n-N_{\rm h}} |  \dots  |  \mathbf{U}^{n+1} \right]  \begin{pmatrix}    c_{N_{\rm h}+1}(\alpha) \\
                                \vdots \\
                                 c_1(\alpha).
\end{pmatrix}.
$$
The effect of approximating the visco-elastic process in the range $[t-T_{\rm h}, t]$ of the displacement filed history is displayed in Fig. \ref{trunc} in which the solution obtained with the full Grünwald-Letnikov approximation is compared with the solution obtained with a truncation with a memory horizon $T_{\rm h}$ for the displacement history. The advantage of truncating the approximation of the displacement history is to reduce the computational cost of the simulation. As shown in Fig. \ref{trunc} this has an effect on the accuracy of the asymptotic solution for large time horizons.

\subsection{Heat transfer in coupled thermo-visco-elastic dynamics}
The numerical setup for this example is taken from \cite{FEAP}. Let $R=[0, L] \times [0, H]$ be a square domain in $\mathbb{R}^2$ occupied in its undeformed configuration
by a visco-elastic material in a state of plane strain. The displacement $u = (u_x , u_y )^{\rm T}$
is such that $u_x$ is zero on the right hand side.
Loading is provided by a transient thermal analysis in which the left side
has an imposed temperature $T_{\rm left} = 1  \ \rm K$ (all the quantities are adimensionalized as in \cite{FEAP}), which is suddenly applied at time zero and
held constant. The other normalized parameters are the Young modulus $E = 100 \ \rm Pa$, the Poisson coefficient
$\nu = 0.4995$,
the thermo-elastic coupling factor $\alpha_T = 0.25$, 
the initial temperature $T_0 = 0 \ \rm K$,
the thermal conductance $k = 10 \ \rm W/ (m^2 K) $,
the heat capiacity $c =1$,
and the density 
$\rho = 0.1 \ \rm Kg/m^3$, see \cite{FEAP}. The governing problem to be solved in $R \times [0, t_{\rm fin}]$ is:

\begin{align}
- \text{div}(\boldsymbol{\sigma}(\mathbf{u}, T)) &=0 \\
\rho c \dfrac{\partial T}{\partial t} +T_0 \beta \text{div} \left(\dfrac{\partial \mathbf{u}}{\partial t} \right) &= k \nabla^2 T.
\end{align}

The thermal stress $\boldsymbol{\sigma}$ is given by:
$$
\boldsymbol{\sigma}= A \mathbf{G} D_{A, \alpha} \boldsymbol{\varepsilon} + \left( K \boldsymbol{\varepsilon} + \beta (T-T_0) \right) \mathbf{I},
$$
where $D_{A, \alpha}$ is the fractional-thermal derivative describing
the thermal relaxation behavior is induced by the heat conduction equation. 
 An Euler backward scheme for the time integration of the thermal problem has been adopted, in which the timestep is $\Delta t = 0.005$. 
Given the temperature $T^n$
at the previous time $t^n$, let $m_k \leq n$ be the current value of the thermal clock. The visco-elastic problem consists in: find the current displacement $\mathbf{u}^{n+1}$, such that
for all test function $\mathbf{v}$ holds:
\begin{equation}
\begin{aligned}
& \int_R \left( (\Delta t)^{-\alpha(T^{m_k})} A(T^{m_k}) \mathbf{G}+ K \mathbf{I} \right) \boldsymbol{\varepsilon}(\mathbf{u}^{n+1}) : \boldsymbol{\varepsilon}(\mathbf{v})   \; \textrm{d} \mathbf{x}
=
\int_R \beta T^n \text{div}(\mathbf{v}) \; \textrm{d} \mathbf{x} -  \\
& \int_R (\Delta t)^{-\alpha(T^{m_k})} A(T^{m_k}) \mathbf{G} \left( \sum^n_{j=m_k} c_{n+2-j}(\alpha(T^{m_k})) (\mathbf{u}^j) \right): \boldsymbol{\varepsilon}(\mathbf{v}) \; \textrm{d} \mathbf{x}. 
\end{aligned}
\end{equation}
Then, given $\mathbf{u}^{n+1}$, solution of the previous problem, the temperature at the current time $T^{n+1}$ is such that, for each test function $s$:
$$
\int_R \rho c \dfrac{\mathbf{u}^{n+1}- \mathbf{u}^n}{\Delta t} s \; \textrm{d} \mathbf{x} 
+
\int_R k \nabla T^{n+1} \cdot \nabla s \; \textrm{d} \mathbf{x} 
+
\int_R \beta T_0 \textrm{div} \left( \dfrac{\mathbf{u}^{n+1}-\mathbf{u}^n}{\Delta t} \right) s \; \textrm{d} \mathbf{x}.
$$
 For the simulation $\mathbb{P}^2$ lagrangian triangular elements have been used to approximate the displacement field and $\mathbb{P}^1$ lagrangian trianguar elements have been used to approximate temperature. 
\subsubsection{Case I: Fractional visco-elastic parameters $A=E$, $\alpha=\{0, 0.1, 0.25 \}$}
\begin{figure}[h!]
\centering
 \includegraphics[width=6cm]{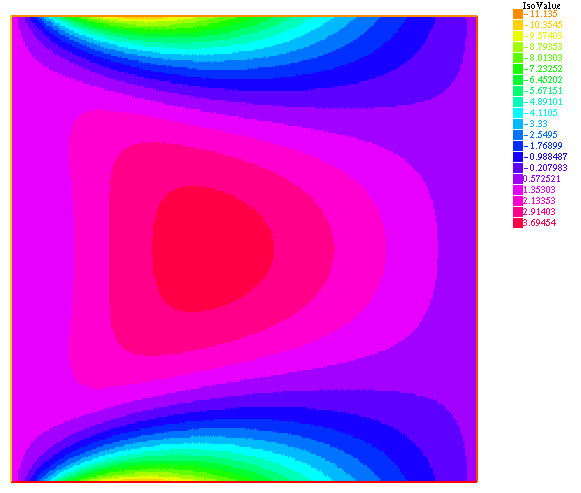} \qquad %te_e_1.png ss1
 \includegraphics[width=6cm]{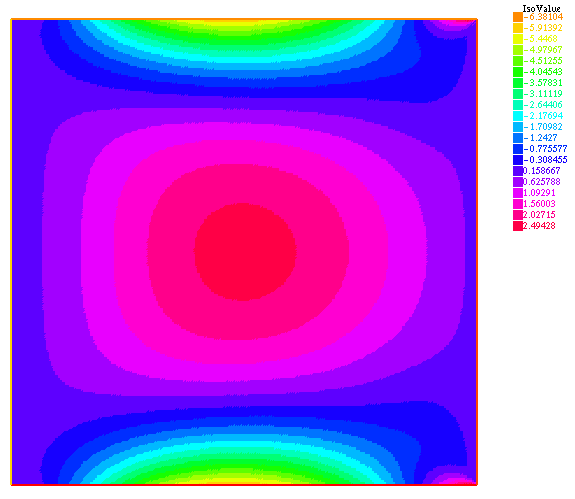} %te_e_fin.png ss2
  \caption{Contour plots of the horizontal component
$\sigma_{xx}$ of the stress at initial and final time ($\alpha=0$).}
 \label{es4}
\end{figure}

In this first example, the material parameters for the fractional derivative are taken as $A=E$ fixed and the fractional exponent is varied as $\alpha=\{0, 0.1, 0.25 \}$. 
In Fig. \ref{es4}, the contour plots of temperature $T(x, y, t)$ and
the horizontal component $\sigma_x (x, y, t)$ of the stress tensor for the initial and
final time  are shown. Temperature is diffusing linearly inside the
region from the left lateral side of the square. 
In Fig. \ref{a0}, the spatial evolution of the temperature profile and thermal stress component $\sigma_{xx}$ are plotted along the central line of the square $\{y=H/2 \}$ for subsequent times in the thermo-elastic case ($\alpha=0$). In Fig. \ref{a0}, the spatial evolution of the thermal stress component $\sigma_{xx}$ along the central line of the square $\{y=H/2 \}$ is shown for subsequent times and for different values of the fractional exponent $\alpha=0.1, 0.25$.

\begin{figure}[h!]
\centering
 \includegraphics[width=6cm]{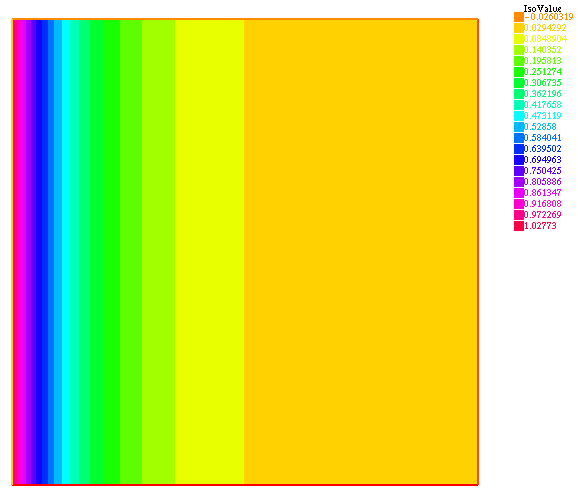} \qquad %te_t_1.png
 \includegraphics[width=6cm]{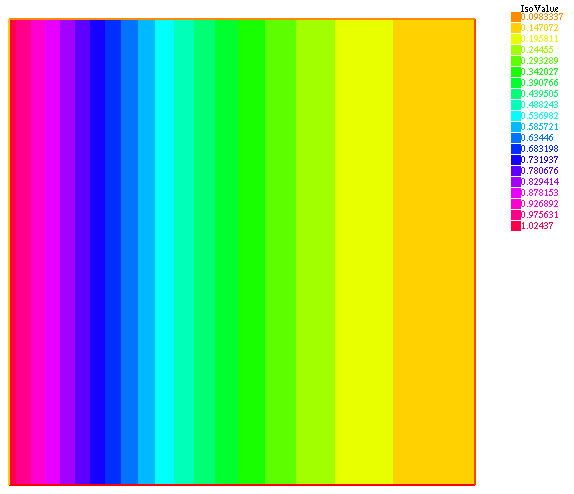} %te_t_fin.png
  \caption{ Contour plots of the temperature $T$ at initial and final time ($\alpha=0$).}
 \label{FIG_1}
\end{figure}

%\begin{figure}[h!]
%\centering
% \includegraphics[width=6.5cm]{thermoela1.png} \qquad %te_t_1.png
% \includegraphics[width=6.5cm]{thermoela2.png} %te_t_fin.png
%  \caption{ Contour plots of the Temperature $T$ at initial and final time.}
% \label{FIG_1}
%\end{figure}

\begin{figure}[h!]
\centering
 \includegraphics[width=6.25cm]{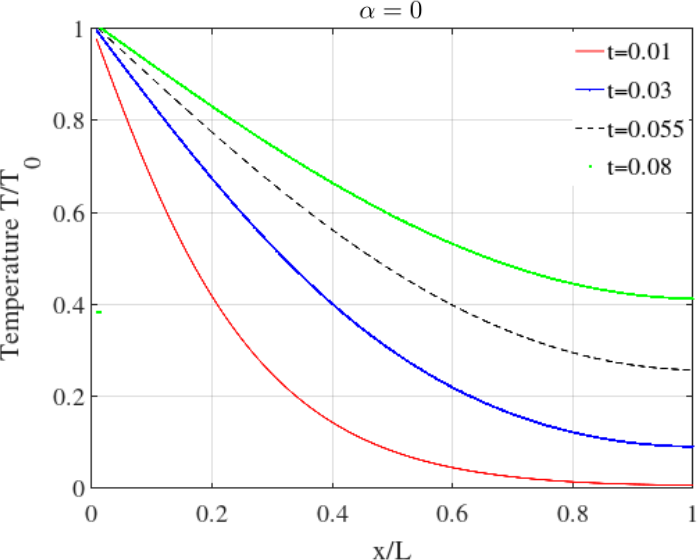} \qquad % a1 te_t_1.png
 \includegraphics[width=6.25cm]{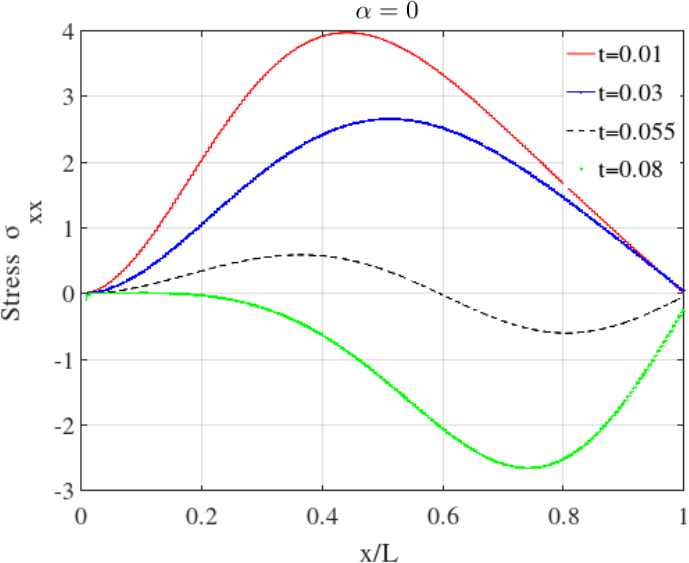}  \\ % a2
  \caption{Numerical results of the spatial evolution of the temperature $T$ (left) and stress component $\sigma_{xx}$ (right) along the line $\{y=H/2 \}$ at different normalized times for the elastic limit $\alpha=0$.}
 \label{a0}
\end{figure}

\begin{figure}[h!]
\centering
\includegraphics[width=6.25cm]{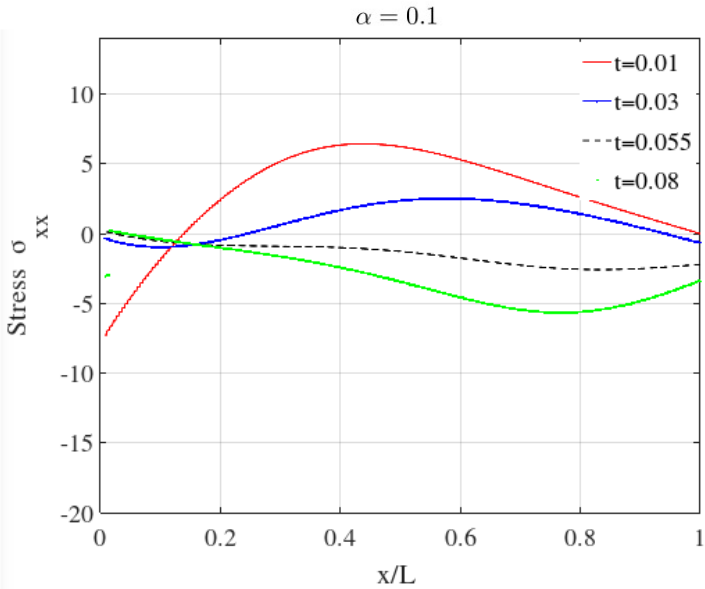} %a1_01.png
\qquad
 %te_t_fin.png
\includegraphics[width=6.25cm]{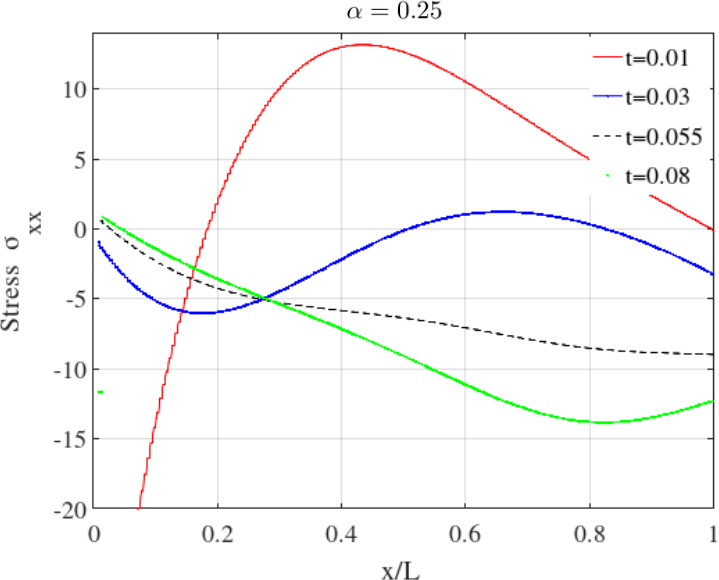} %a1_025.png
  \caption{ Numerical simulation of the the spatial evolution of the stress component $\sigma_{xx}$ along the line $\{y=H/2 \}$ at different normalized times for different values of the fractional exponent $\alpha=\{0.1, 0.25 \}$.}
 \label{a1}
\end{figure}

\newpage
\subsubsection{Case II: Fractional visco-elastic parameters $A$, $\alpha$ temperature dependent}
In this example, the material functions $A(T)$ and $\alpha(T)$ are considered as functions of temperature $T$ as:

\begin{equation}\label{eq:last}
A(T)=\begin{cases}
       10E \quad & \text{if} \ T \leq T^*, \\
       0.1E \quad & \text{if} \ T >    T^*
     \end{cases},
     \qquad 
\alpha(T)=\begin{cases}
           0.05    \quad & \text{if} \ T \leq T^*, \\                                           0.25  \quad   & \text{if} \ T >    T^*
          \end{cases}. 
\end{equation}
where $T^*=0.5$.

\begin{figure}[h!]
\centering
 \includegraphics[width=6.25cm]{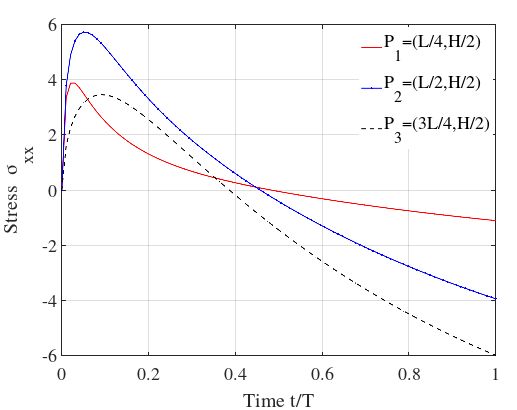} \qquad %te_e_1.png ss1
 \includegraphics[width=6.25cm]{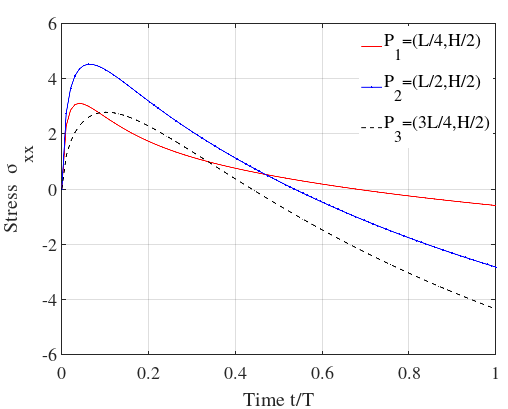} %te_e_fin.png ss2
  \caption{Numerical solution of the temporal evolution of the horizontal component
$\sigma_{xx}$ of the stress on different points $P_1=(L/4,H/2)$, $P_2(L/2,H/2)$ and $P_3=(3L/4,H/2)$ of the square domain $R=[0, L] \times [0,H]$ for the thermo-elastic case $\alpha=0, A=E$ (left) and for $\alpha$ and $A$ as in Eq. \eqref{eq:last} (right).}
 \label{eslast}
\end{figure}

The temporal evolution of the stress component $\sigma_{xx}$ is shown in Fig. \ref{eslast} for three points in the square domain $R=[0, L] \times [0, H]$, respectively $P_1=(L/4,H/2)$, $P_2(L/2,H/2)$ and $P_3=(3L/4,H/2)$ for the cases $\alpha=0, A=E$ (thermo-elastic case) and for $\alpha$ and $A$ as in Eq. \eqref{eq:last}.

\clearpage
\newpage
\section{Conclusions}
A novel finite element computational framework for the simulation of coupled thermo-visco-elasticity problems in thermo-rehologically complex materials with memory has been proposed. The visco-elastic constitutive law is based on fractional calculus. Fractional calculus and the theory of Mittag-Leffler special functions allow an accurate
description of the time relaxation behaviour of the material. The fractional parameters of the model are assumed to be temperature dependent. An internal material clock has therefore been introduced to
model the thermo-rheological complexity of materials in which the classical time-temperature superposition
principle does not apply. The model is able to represent  the phase transition experienced by the relaxation modulus as a function of temperature. The numerical treatment of the fractional derivative 
has been done via the Grünwald-Letnikov approximation which leads to an
additional load vector which represents the memory of the material in the discretized system of equations resulting from the finite element formulation.
\\
The poposed model has been implemented in the finite element software FreeFem++ \cite{FF++} and has been used to solve numerically various test cases, namely: (i) the free vibrations of a visco-elastic rod; (ii) creep tests; (iii) heat transfer in coupled thermo-visco-elastic dynamics. The proposed model has been validated against creep and relaxation tests conducted at different temperatures on EVA, a real visco-elastic material which is a polymer used as encapsulant for solar cells. With the present computational tool available, further
developments may regard the simulation of thermo-visco-elastic phenomena observed in experiments, opening the possibility to accurately simulate with the finite element method any arbitrary rheologically complex material for any application and problem geometry.
\\
\\
\textbf{Acknowledgements} \\ \\
The support from the Italian Ministry of Education, University and Research (MIUR) to the Research Project of Relevant National Interest (PRIN 2017) "XFAST-SIMS: Extra-fast and accurate simulation of complex structural systems" (CUP: D68D19001260001) is gratefully acknowledged.

\end{document}